\documentstyle[psfig]{l-aa}

\def\ros{{\sl ROSAT}}
\def\etal{{et\,al. }}

\def\degs{\ifmmode ^{\circ}\else$^{\circ}$\fi}
\def\amin{\ifmmode ^{\prime}\else$^{\prime}$\fi}
\def\asec{\ifmmode ^{\prime\prime}\else$^{\prime\prime}$\fi}
\def\fd{\hbox{$.\!\!^{\rm d}$}}            
\def\fss{\hbox{$.\!\!^{\rm s}$}}        
\def\farcs{\hbox{$.\!\!^{\prime\prime}$}}  
\def\h{$^{\rm h}$}\def\m{$^{\rm m}$}

\newbox\grsign \setbox\grsign=\hbox{$>$}
\newdimen\grdimen \grdimen=\ht\grsign
\newbox\laxbox \newbox\gaxbox
\setbox\gaxbox=\hbox{\raise.5ex\hbox{$>$}\llap
     {\lower.5ex\hbox{$\sim$}}}\ht1=\grdimen\dp1=0pt
\setbox\laxbox=\hbox{\raise.5ex\hbox{$<$}\llap
     {\lower.5ex\hbox{$\sim$}}}\ht2=\grdimen\dp2=0pt

\def\lax{\mathrel{\copy\laxbox}}
\def\rxj{RX\,J0719.2+6557}

\begin{document}

   \thesaurus{06         
              (02.01.2;  
               08.02.2;  
               08.09.2;  
               08.13.1;  
               08.14.2;  
               13.25.5)} 

   \title{RX J0719.2+6557: A new eclipsing polar}

   \author{G.H. Tovmassian\inst{1}
   \and J. Greiner\inst{2}\thanks{Present address: Astrophysical Institute
        Potsdam, An der Sternwarte 16, 14482 Potsdam, Germany}
   \and F.-J. Zickgraf\inst{2}\thanks{Present address: Observatoire de
         Strasbourg, 11 rue de l'Universit\'e, F-67000 Strasbourg, France}
   \and P. Kroll \inst{3}
   \and J. Krautter \inst{4}
   \and I. Thiering \inst{4}
   \and S.V. Zharykov \inst{5}
   \and A. Serrano \inst{6}
          }

   \offprints{G.H. Tovmassian, gag@bufadora.astrosen.unam.mx}

   \institute{Observatorio Astronomico Nacional, Instituto de Astronom\'{\i}a, 
       UNAM, AP 877, 22860, Ensenada, B.C., M\'{e}xico 
   \and Max-Planck-Institut f\"ur extraterrestrische Physik,
             85740 Garching, Germany
   \and Sternwarte Sonneberg, 96515 Sonneberg, Germany
   \and Landessternwarte K\"onigstuhl, 69117 Heidelberg, Germany
   \and Special Astronomical Observatory, 357147 Nizhnij Arkhyz, Russia
   \and Instituto Nacional de Astrof\'{\i}sica Optica y Electr\'onica, 
     AP 51 y 216, Puebla, Pue., Mexico    }

   \date{Received 31 December 1996 / Accepted 1 August 1997}

   \maketitle

   \begin{abstract}

A new magnetic cataclysmic variable is identified as the counterpart of the
X-ray source RX J0719.2+6557 which has been discovered during the ROSAT 
all-sky survey. Optical spectroscopy of the $\approx 17^{\rm m}$ object 
revealed a pattern of Balmer and strong He~II emission lines which are
characteristic for cataclysmic variables.  The emission lines
show radial velocity variations with a period of 98.2 min with no significant
phase shifts among them.
This coincides with the period of deep eclipses (up to 4 mag) in the
photometric light curve of the system. The phase of the
eclipse relative to the spectroscopic phase, and its structure indicates that 
the dominant source of emission is located on the stream of accreting matter, 
which is eclipsed by the secondary companion. The emission lines bear evidence
 of a weaker component,  most probably the contribution from the heated side 
of the secondary star. These features define this object as a probable 
polar in a high state in which the secondary is irradiated by the
X-rays originating from magnetically driven accretion onto the white dwarf.
The near-infrared spectroscopy revealed some unusual, strong emission features
at 8200 \AA\ and 8660 \AA\ possibly originating on the heated side of the 
secondary.

   \end{abstract}

\section{Introduction}

Within a program devoted to the optical identification of a complete
sample of northern ROSAT all-sky survey (RASS) X-ray sources
we identified a new cataclysmic variable.
The identification project is a collaboration
of the Max-Planck-Institut f\"ur extraterrestrische Physik, Garching, 
Germany, with the the Landessternwarte Heidelberg (LSW), Germany, and the 
Instituto Nacional de Astrofisica, Optica y Electronica (INAOE), Puebla, 
Tonantzintla, Mexico. A detailed description of the 
project is given by Zickgraf et al. (1997). 

For the study areas the 
individual scan strips of the  all-sky survey were merged to 
produce a final data base comprising about 1600 X-ray sources with a detection 
likelihood larger than 10 corresponding to X-ray detection limits of 
$\sim 0.003$ cts\,s$^{-1}$ in area near the North Ecliptic Pole and 
$\sim 0.01$ cts\,s$^{-1}$ in all 
other areas. Since October 1991 nearly 800 X-ray sources have been observed
within the identification program.
Here we report the identification and detailed follow-up observations of the
ROSAT all-sky survey  X-ray source RX\,J0719.2+6557 (= 1RXS J071913.4+655734),
first results of which have been reported already earlier (Tovmassian \etal 
1997).

Cataclysmic variables (CVs) are close binary systems, in which a compact 
white dwarf (WD) primary accretes matter from a Roche-lobe filling late-type 
main-sequence secondary component. 
Polars are a subclass of cataclysmic variables where the magnetic field of 
the white dwarf is strong enough to channel the accretion along the
magnetic field lines, thus preventing the formation of an accretion disc 
and synchronizing the rotation of the white dwarf with the orbital period 
(Warner 1995). A hot spot (or accretion spot) which is formed as a 
result of a shock of accreting matter before hitting the white dwarf 
gives rise to hard X-ray radiation (often described with a bremsstrahlung 
model). 

\begin{table*}[th]
\vspace{-0.25cm}
\caption{Log of Optical Observations}
\vspace{-0.2cm}
\begin{tabular}{ccccrrl}
      \noalign{\smallskip}
      \hline
      \noalign{\smallskip}
Date & JD & Telescope + Equip. & Filter       & Duration & Exposure & ~~Site \\
     &    &                    & Wavelength   &  (min.)~~  & (sec.)~~ &  \\ 
 \noalign{\smallskip}
 \hline
 \noalign{\smallskip}
 1996 April 07   & 2450180 & 2.1m, B\&Ch sp. & 3600--5400 & 60~~ & 1200 & SPM\\
 1996 April 08   & 2450181 & 2.1m, B\&Ch sp. & 3600--5400 & 180~~ & 600 & SPM\\
 1996 April 09   & 2450182 & 2.1m, B\&Ch sp. & 6000--9000 & 80~~ & 600 & SPM\\
 1996 April 10   & 2450183 & 2.1m, B\&Ch sp. & 3600--5400 & 80~~ & 1200 & SPM\\
 1996 April 15    & 2450189 & 0.6m, CCD     & B & 200~~ & 60,120 & Sonneberg\\
 1996 April 16    & 2450190 & 0.6m, CCD     & B & 180~~ & 120 & Sonneberg\\
 1996 April 17    & 2450191 & 0.6m, CCD     & B &  70~~ &  90 & Sonneberg\\
 1996 August 8/9  & 2450304 & 0.6m, CCD     & R & 100~~ & 120 & Sonneberg\\
1996 September 3/4 &2450330 & 0.6m, CCD     & R & 110~~ &  90 & Sonneberg\\
 1996 October 08  & 2450366 & 1.0m, CCD     & B & 135~~ & 300 & Zelenchuk \\
 1996 December 04 & 2450422 & 0.6m, CCD   & B, R & 220~~ & 90,120 & Sonneberg\\
 1996 December 10 & 2450428 & 0.6m, CCD   & B, R & 380~~ & 90 & Sonneberg\\
       \noalign{\smallskip}
      \hline
    \end{tabular}
   \label{log}
   \end{table*}

\section{X-ray observations}

The location of \rxj\ was scanned in the RASS during September 22--24, 1990
for a total exposure time of 570 sec. \rxj\ is found as a source with 
a total of 66 photons, which corresponds to a vignetting corrected
countrate of 0.16 cts/s in the ROSAT PSPC. 
No strong variability in the X-ray intensity is seen at this statistics,
in particular also no orbital variations.
The spectrum as derived from these 66 photons is rather hard, extending
up to 2.4 keV (the upper bound of the PSPC). The standard hardness ratio
(HR1 is the normalized count difference
(N$_{\rm 52-201}$ -- N$_{\rm 11-41}$)/(N$_{\rm 11-41}$ + N$_{\rm 52-201}$),
where N$_{\rm a-b}$ denotes the number of counts in the PSPC between
channel a  and channel b, and HR2 is similarly defined as
(N$_{\rm 91-200}$ -- N$_{\rm 50-90}$)/N$_{\rm 50-200}$ with the count 
number divided by hundred corresponding roughly to the energy in keV)
values are HR1 = 0.48$\pm$0.12 and HR2 = 0.50$\pm$0.12.
A thermal bremsstrahlung model
gives a reasonable fit for a rather wide range of temperatures (2-15 keV).
For a fixed temperature of kT=10 keV the best fit absorbing column is
1.6$\times$10$^{20}$ cm$^{-2}$, and the unabsorbed flux is 
2$\times$10$^{-12}$ erg/cm$^2$/s in the 0.1--2.4 keV range. This is a lower 
limit because any, even very weak, soft component as 
usually found in polars would require a higher absorbing column to be 
compatible with the observed X-ray spectrum. \rxj\ has not 
been covered by any (serendipitous) ROSAT pointing until the time of this 
writing, so that the X-ray parameters cannot be improved further.

Given the rather low absorbing column (which varies by only 30\% for
models with different temperature) as compared to the total galactic column in 
the direction of \rxj\, of 4.6$\times$10$^{20}$ cm$^{-2}$ (Dickey \& Lockman 
1990) the distance of \rxj\ is possibly only about 100 pc. Again, this is a 
lower limit in the sense that any (expected, but due to statistics and
background contamination not detectable) soft blackbody-like component
would result in an increase of the absorbing column and thus an increased
distance. At this distance the (lower limit of the) phase-averaged, 
unabsorbed X-ray luminosity is 
2.5$\times$10$^{30} (D/100 {\rm pc})^2 $ erg/s in the 0.1--2.4 keV range.
   
\section{Results of optical observations}

\subsection{Observational details}

The original identification observations were carried out at the 2.15\,m 
telescope of the Guillermo Haro Observatory which is operated by INAOE and is 
located near Cananea, Sonora, Mexico. For the purpose of the identification
project the LSW has constructed an efficient faint object spectrograph 
($LFOSC$). It allows to carry out direct CCD imaging, filter photometry,
and, in particular, multi-object spectroscopy by using interchangeable 
hole masks with circular holes in the focal plane of the telescope 
(Zickgraf \etal 1997).
Grisms follow in the parallel beam section of the focal reducer. The hole 
masks are produced from the CCD frames of the direct images with a computer 
controlled drilling device. Two grisms are available, giving a resolution 
of 13 {\AA} and 18 {\AA}, respectively. 
First direct imaging was carried out on January 24, 1993, revealing the 
presence of four candidates brighter for the optical counterpart than $R = 
23^{\rm m}$  in the error circle of RX\,J0719.2+6557. Multi-object 
spectroscopy with a resolution of 18\AA\ of the candidates was performed in 
January 25, 1993. Inspection of the spectra showed the optically brightest 
candidate to be an emission line object with
spectral characteristics typical for cataclysmic variables. 

\begin{table}[ht]
\vspace{-0.45cm}
\caption{Observed minima (eclipses) in the B band}
\vspace{-0.2cm}
\begin{tabular}{clrl}
      \noalign{\smallskip}
      \hline
      \noalign{\smallskip}
Date & Min. (HJD) & E & (O--C) \\ 
 \noalign{\smallskip}
 \hline
 \noalign{\smallskip}
 1996 April 15  & 2450189.46324 &  0 & \,--0.00002 \\
 1996 April 16  & 2450190.35017 & 13 &  +0.00023 \\
 1996 April 16  & 2450190.41861 & 14 &  +0.00047 \\
 1996 April 17  & 2450191.44190 & 29 &  +0.00066 \\
 1996 October 08 & 2450366.52778 & 2596 & --0.00083 \\
 1996 December 04 & 2450422.39000 & 3415 & --0.00014 \\
 1996 December 04 & 2450422.45704 & 3416 & --0.00131 \\
 1996 December 10 & 2450428.52597 & 3505 & --0.00282 \\
 1996 December 11 & 2450428.59522 & 3506 & --0.00178 \\
 1996 December 11 & 2450428.66326 & 3507 & --0.00195 \\
      \noalign{\smallskip}
      \hline
    \end{tabular}
   \label{min}
   \end{table}

Further detailed study of the emission line object, and its final 
identification as a new polar, 
was performed at the 2.1\,m telescope of the Observatorio Astron\'omico 
Nacional (OAN) de San Pedro M\'artir,
Mexico in early April 1996. The Boller \& Chivens spectrograph was used to 
obtain spectra of the object. Two wavelength ranges were observed. 
A 600 l/mm grating was used to obtain spectra in the $\lambda\lambda 
3600 - 5400 \AA$ range. In combination with a 2\arcsec\, slit 
a   4 $\AA$ FWHM resolution is reached. The $\lambda\lambda 6000 - 9000 \AA$ 
range was covered with 6 $\AA$ resolution using a 400 l/mm grating. 
Exposure times of 600 sec. 
were chosen as a compromise between high temporal resolution in order to 
derive the orbital period of the system and well exposed spectra.
The seeing during the first three nights was about 1\farcs5, while in the last 
night it was 2\farcs3-2\farcs5, and the air was contaminated by dust 
brought up by storm. 
In total, we obtained 23 spectra in the blue part, distributed over three 
nights. Additional 8 spectra were obtained in the red side of the spectrum 
during one night. The standard stars Feige\,34 and HZ\,44 were observed each 
night for flux calibration. Comparison spectra were taken every 20 minutes. 
They were used for wavelength calibration precise up to $0.3 \AA$ and actual 
dispersion solutions were assigned to the object spectra based upon  the 
julian date. The spectra were reduced using standard routines in 
the IRAF package. Spectrum extraction was performed according to the optimal 
extraction method as described by Horne (1986). 

Extensiv optical photometry was first performed on three nights in 
1996 April (15/16 and 
16/17 and 17/18, some ten days after the spectroscopic observations)
at the Sonneberg Observatory 600/1800 mm reflector, equipped with a
385$\times$578 pixel EEV CCD. Exposure times were 60
to 120 sec., and a Johnson B filter was used.
Further photometry was acquired in August and September 1996 in the R band,
and in December 1996 with alternating
exposures using Johnson B and R filters, respectively.

One photometric run was also obtained at the Special Astrophysical Observatory
(Zelenchuk) 1 m telescope over more than two hours with 5 min. exposures each 
in the B band. 
A log of all optical observations is summarized in Tab. \ref{log}.

\begin{figure}[t]
      \vbox{\psfig{figure=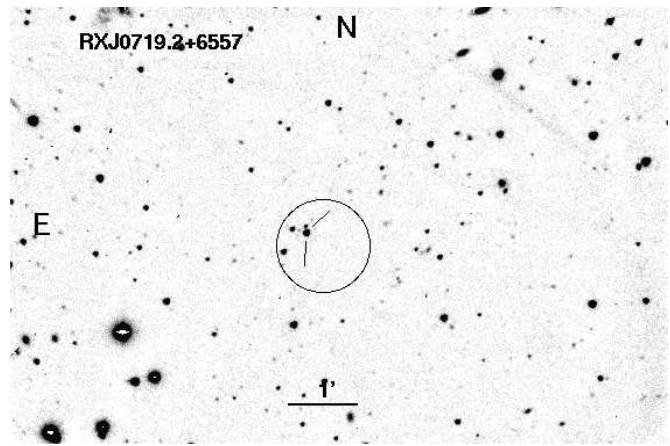,width=8.8cm,%
          bbllx=2.4cm,bblly=16.1cm,bburx=18.1cm,bbury=26.5cm,clip=}}\par
      \caption[fchart]{$R$ band image of the field around the X-ray source
          RX\,J0719.2+6557 (centroid position with a 40\arcsec\, (3$\sigma$)
          error circle). The cataclysmic variable is marked by two dashes.}
      \label{fc}
\end{figure}

\subsection{Identification and position}

The sequence of spectra of the emission line object taken with the 2.1 m 
telescope shows strong emission lines of the Balmer series, He~I and He~II
on top of a blue continuum. Though no polarimetric measurements were
performed, the relative line strengths are characteristic for magnetic 
cataclysmic variables (CVs).

We measure the position of the optical counterpart of \rxj\ as (equinox 2000.0)
R.A. = 07\h19\m14\fss0, Dec. = 65\degr57\arcmin48\arcsec ($\pm$1\arcsec).
Fig. \ref{fc} shows a finding chart with the magnetic cataclysmic variable 
marked.

\subsection{Eclipse lightcurve}

\begin{figure}
      \vbox{\psfig{figure=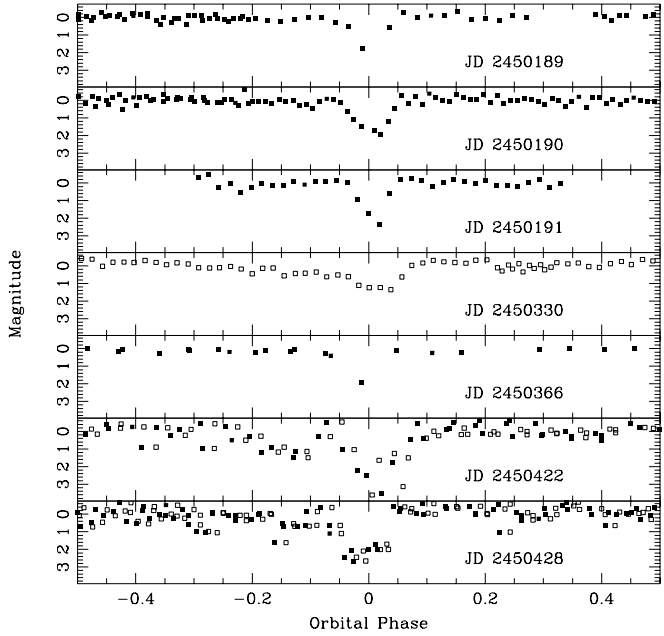,width=8.8cm,%
          bbllx=0.6cm,bblly=5.7cm,bburx=19.9cm,bbury=24.5cm,clip=}}\par
  \vspace{-0.25cm}
 \caption[lcurve]{Light curves of RX\,J0719.2+6557 folded with the
      orbital period. Filled (open) symbols denote B (R) band 
      measurements. The errors are smaller than the symbol size.
      Note the varying width of the flux depression before the eclipse
      as well as the varying eclipse depth.
            }
\label{rbfold}
\end{figure}

Optical photometry clearly shows an eclipsing light curve  (Fig. \ref{rbfold}).
The eclipse FWHM is about 5~min, and the eclipse depth is strongly
variable reaching a maximum amplitude of nearly 4 mag in the blue band. 
The short ingress and egress durations suggest that the emission region
is compact, and we associate it with the hot spot on the accretion stream 
toward the white dwarf.
Our temporal resolution (typically 2 min.) does not allow to investigate 
possible persistent structures in the eclipse ingress or egress.
The large amplitude of the eclipses implies that this compact emission region
is by far the dominant light source in this binary system. 

\begin{figure}
    \vbox{\psfig{figure=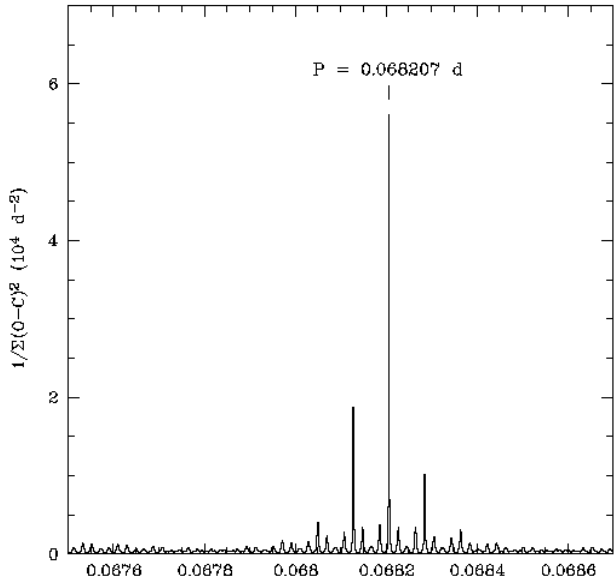,width=8.8cm,%
          bbllx=0.7cm,bblly=5.8cm,bburx=20.cm,bbury=24.5cm,clip=}}\par
 \vspace{-0.25cm}
 \caption[lcurve]{Result of our period analysis (see text). The value given at 
                  the y-axis is the inverted sum of the squared (O-C)-values 
                  resulting from the eclipse timings observed. The period 
                  derived from our linear regression is marked.}
\label{ps}
\end{figure}

The eclipses are regular and stable over months.
The eclipse moments were measured as a center of a gaussian fit to the eclipse
profile in the B band.
We searched for the most probable period using a least squares method applied 
to the timings of the eclipse minima given in Tab. \ref{min}. For a fixed 
$T_{0} =2450189.46324$ we calculated the phasing of all timings for a range 
of trial periods. The inverted sum of the squared (O--C)  values resulting 
from that, is plotted as a function of the trial period in Fig. \ref{ps}. 
It shows a prominent peak at $P = 0.068207 \pm 0.000019$ days, which we 
adopted as the  eclipse period.
We derive the following ephemeris:

$$ T_{\rm ecl} = {\rm HJD} 2450189.46326 + 0\fd068207[19] \times {\rm E}  $$ 

\noindent In the following $\phi_{\rm orb}$ refers to this ephemeris.

\subsection{Orbital modulation}

\begin{figure}
      \vbox{\psfig{figure=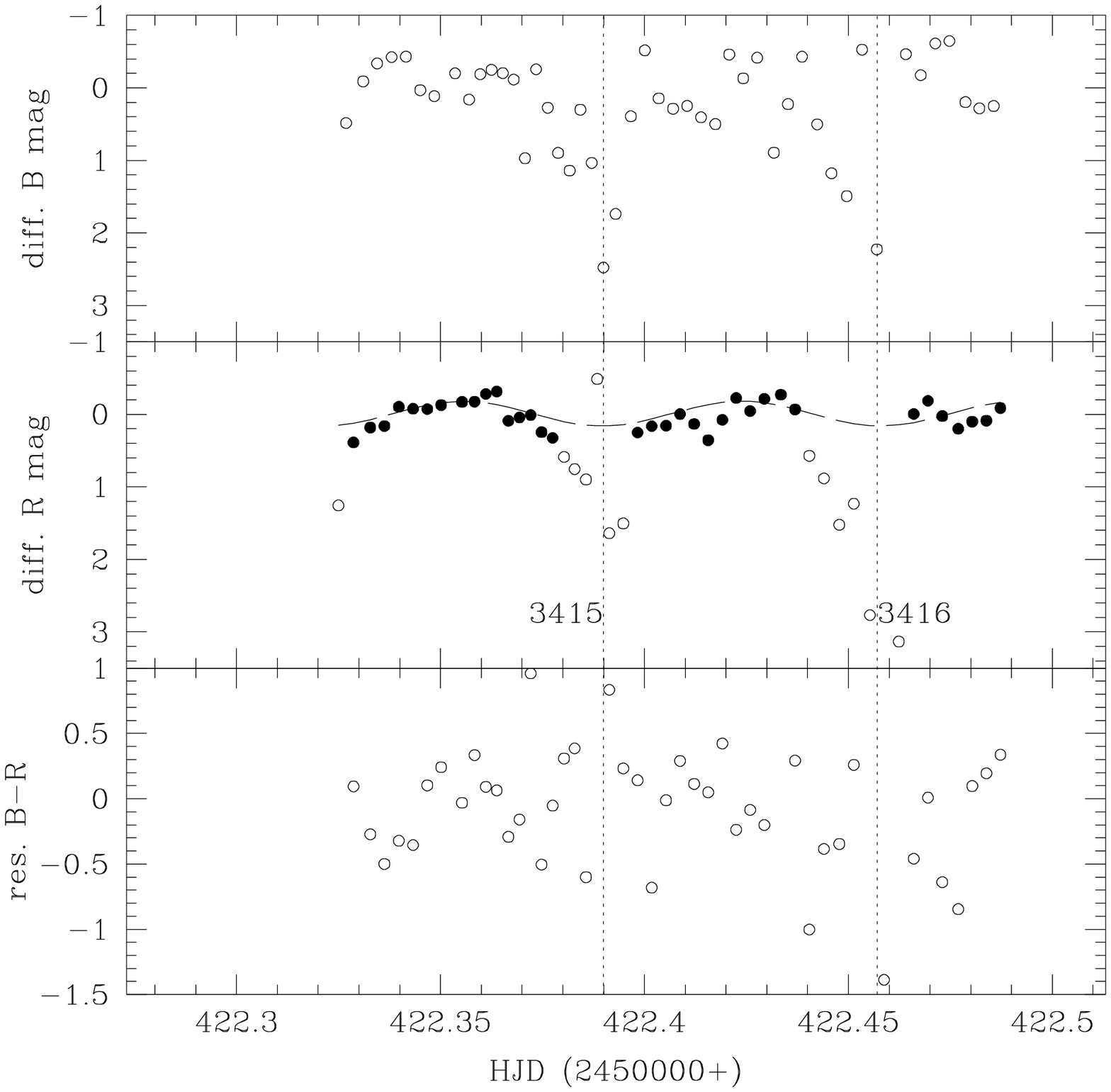,width=8.8cm,%
          bbllx=0.5cm,bblly=5.6cm,bburx=21.5cm,bbury=26.1cm,clip=}}\par
 \vspace{-0.5cm}
      \vbox{\psfig{figure=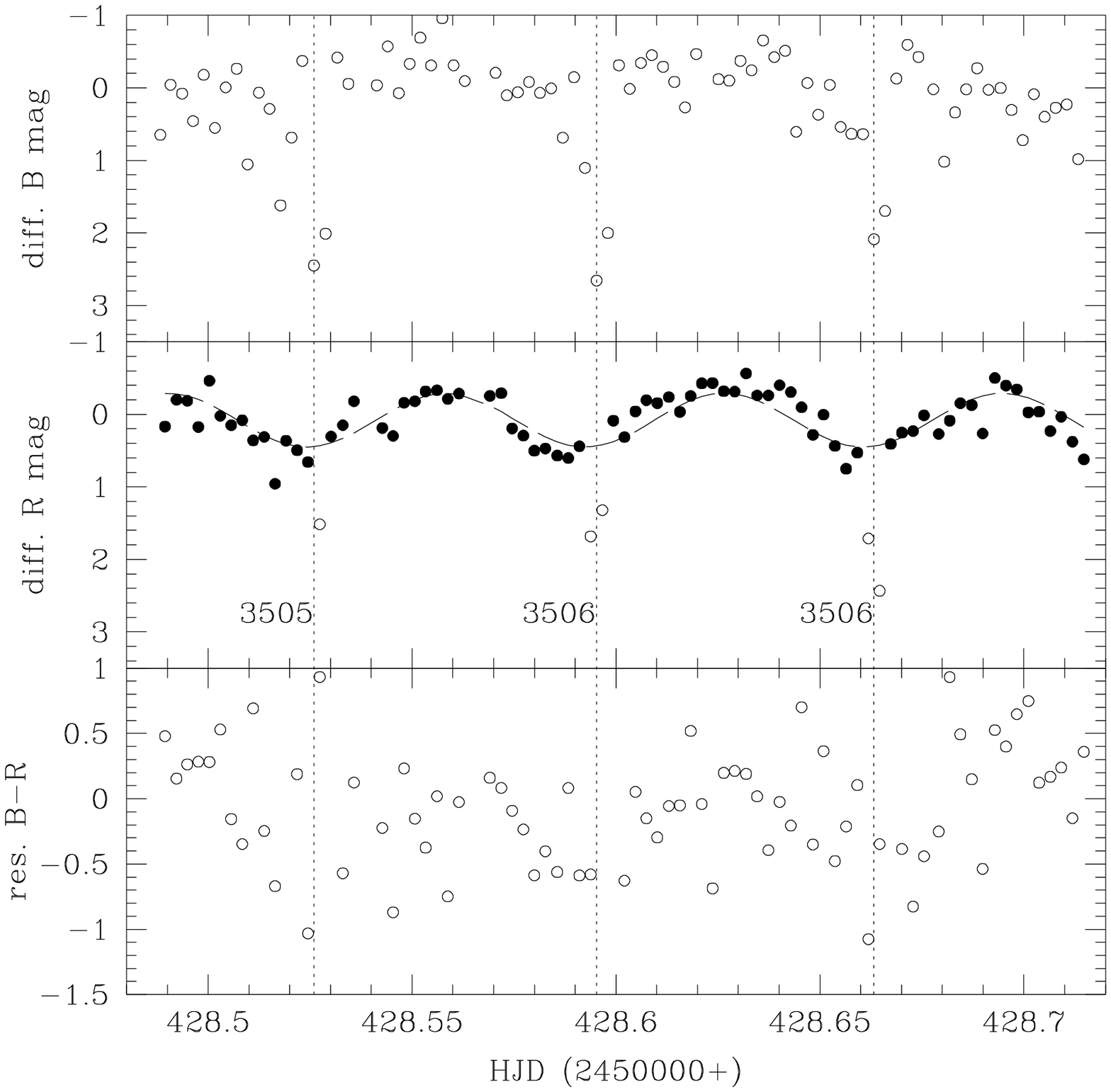,width=8.8cm,%
          bbllx=0.5cm,bblly=5.6cm,bburx=21.5cm,bbury=26.1cm,clip=}}\par
 \vspace{-0.15cm}
 \caption[lcurve]{Two sequences of alternating B and R exposures of
          RX\,J0719.2+6557 obtained on Dec. 4/5 and Dec. 10/11, respectively.
          The error of each data point is about the symbol size.
          The vertical dotted lines mark the times of the mid-eclipse 
          according to
          the photometric ephemeris, and numbers in the R band panels
          denote the cycle number (see also Tab. \ref{min}). The filled circles
          in the R band panels were used to fit a sin curve, while
          open circles (during  eclipse) were omitted.
          The relative colour B--R has been computed by equating the mean
          between two B measurements with the simultaneous R measurement.
            }
\label{brlc}
\end{figure}

In addition to the eclipse profile, another noteworthy feature of the light 
curves is the strong  modulation seen outside the eclipse. This feature is 
not strongly persistent, and most prominently seen in the R band (see Fig.
\ref{brlc}). Though one would expect the light curve to be more complicated, 
the R band variations could be successfully fitted by a sinusoid,
if we ignore the  points within the eclipses 
(Fig. \ref{brlc}). We think that the R band variations are due to the 
varying aspect of the X-ray heated side of the secondary although the 
directionality of optical synchrotron emission is a viable alternative.
If we proceed with this interpretation, it is worth to note  that the
eclipse occurs with a slight delay  relative to the secondary inferior 
conjunction, estimated as a minima on a fitted sin curve. It could be 
attributed to the fact, that the hot spot actually is located at some 
height (i.e. on the accretion stream)  or that the heated spot on the surface 
of the secondary is not symmetric. However, the latter is less probable and 
we will see in the next paragraph that spectroscopic observations also favour 
the first interpretation.

\subsection{Spectral and radial velocity variations}

\begin{figure*}
    \vbox{\psfig{figure=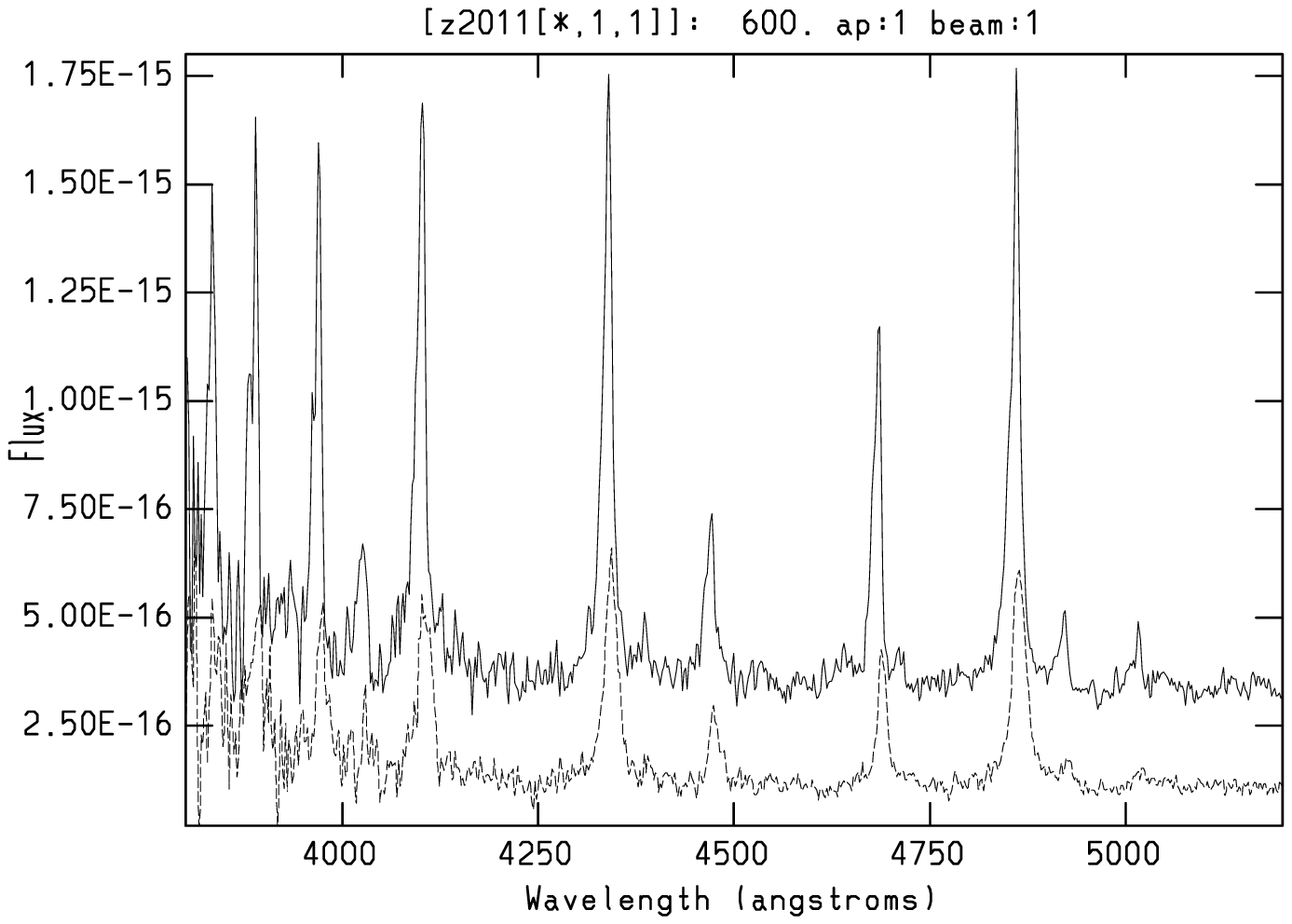,width=8.5cm,%
          bbllx=2.7cm,bblly=9.0cm,bburx=16.8cm,bbury=18.6cm,clip=}}\par
     \vspace*{-5.8cm}\hspace*{8.8cm}
      \vbox{\psfig{figure=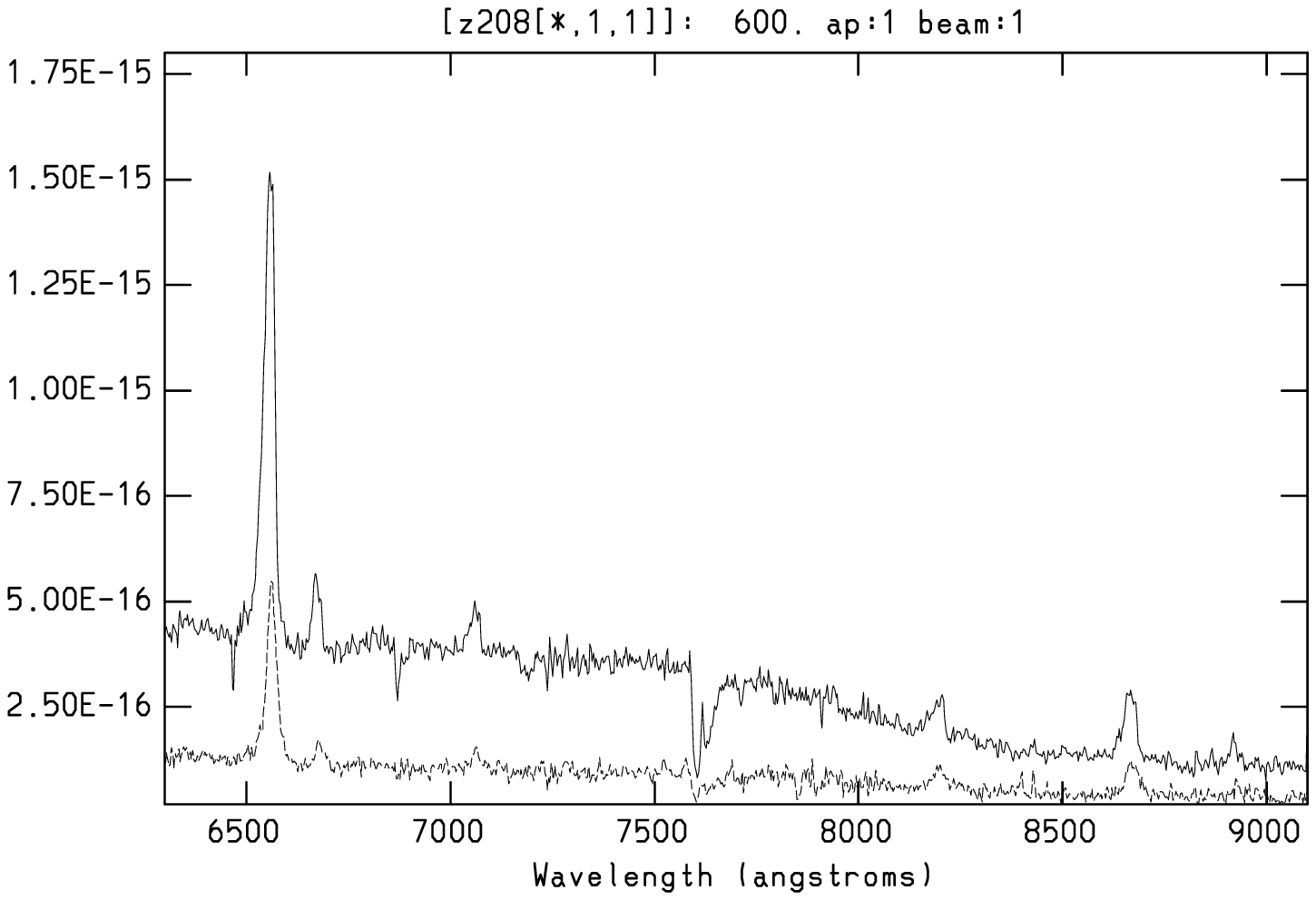,width=9.1cm,%
          bbllx=2.4cm,bblly=9.cm,bburx=17.3cm,bbury=18.6cm,clip=}}\par
 \caption[spectra]{The spectra of \rxj\, in the blue (left panel) and red 
     (right). The solid line represents the spectrum out 
     of eclipse, the dashed line is the spectrum during eclipse.   
     The flux unit is in erg/cm$^2$/s/\AA.
            }
\label{sp}
\end{figure*}

Flux-calibrated spectra of RX J0719.2+6557 in both spectral ranges during 
an eclipse and a phase interval opposite to the eclipse ($\phi_{orb}\simeq 
0.5$) are shown
in Fig. \ref{sp}. The spectra are from the nights with the better seeing.
Although the exposure times are as long as the full duration of the eclipse 
and not exactly centered on it, significant variations between the in- and
out-of-eclipse spectra are observed, in particular concerning the line 
strengths.

\begin{figure}
      \vbox{\psfig{figure=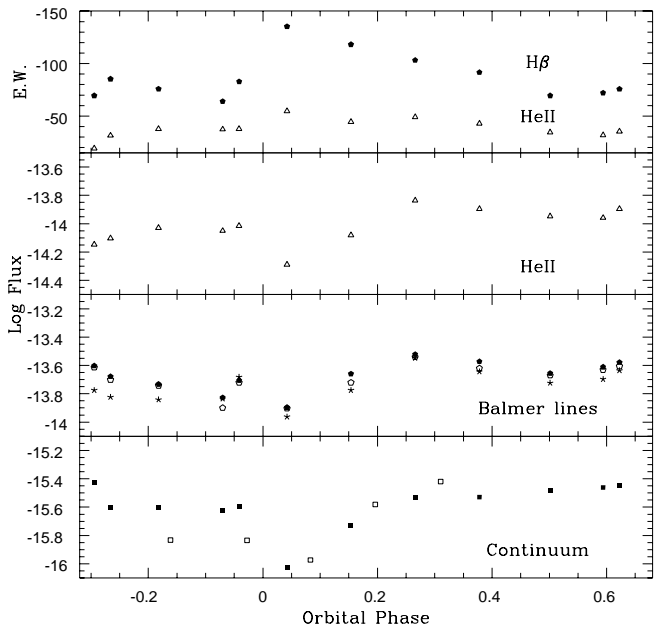,width=8.7cm,%
          bbllx=0.5cm,bblly=5.5cm,bburx=19.9cm,bbury=24.5cm,clip=}}\par
  \vspace{-0.25cm}
 \caption[flew]{Orbital variation of the continuum fluxes
   and major lines. The continuum measurements around $\lambda~4700~\AA$ 
  (filled symbols) and $\lambda~6500~\AA$ (open symbols) are shown in the 
  lower panel. The H${\beta}$ (filled), H${\gamma}$ (open) and 
  H${\delta}$ (starlike symbols) fluxes are in the 
  next panel. In the third panel the fluxes of He~II are displayed. The top 
  panel shows the variation of the equivalent widths of H${\beta}$ and He~II. 
   Flux units are in erg/cm$^2$/s/\AA\, and the equivalent width EW in \AA.
   Because of the time resolution of only 7 min. ($\approx$0.07 phase units)
  achieved in the spectroscopic observations the minima do not exactly match 
  the photometric (orbital) eclipse phase.
            }
\label{fl}
\end{figure}

The mean values of major emission lines and their equivalent widths out of 
eclipse are presented in Tab. \ref{emlns}. The dependence of these values on 
orbital phases  are presented in Fig. \ref{fl}. The continuum fluxes mark the 
eclipse clearly in the lower panel. The Balmer lines show a smaller drop of 
flux strength relative to He~II, while the equivalent width of H$\beta$ 
exhibits a large 
increase during eclipse in contrast to He~II. This implies that the eclipse 
affects the continuum and He~II more than Balmer lines.

\begin{table}
\vspace{-0.25cm}
\caption{Flux and Equivalent Width of Emission Lines}
\vspace{-0.2cm}
\begin{tabular}{lcll}
      \noalign{\smallskip}
      \hline
      \noalign{\smallskip}
Emission Line & Log Flux              & ~~E.W.      & FWHM \\ 
              & ($erg/cm^2/s/\AA$)    & ~~($\AA$)   & ~~($\AA$)  \\
 \noalign{\smallskip}
 \hline
 \noalign{\smallskip}
 H$\delta$  & $-13.70 \pm 0.10 $  & ~$-55 \pm 8$ &  \\
 H$\gamma$  & $-13.65 \pm 0.08 $  & ~$-67 \pm 10$ &  \\
 H$\beta$   & $-13.60 \pm 0.08 $  & ~$-72 \pm 10$ &  $20 \pm 2$ \\
 H$\alpha$  & $-13.50 \pm 0.15 $  & $-101 \pm 14$ &  $30 \pm 3$ \\
 He I 4471  & $-14.12 \pm 0.10 $  & ~$-21 \pm 7$ &  \\
 He~II 4686 & $-14.10 \pm 0.10 $  & ~$-37 \pm 10$ & $14 \pm 4$ \\
       \noalign{\smallskip}
      \hline
    \end{tabular}
   \label{emlns}
   \end{table}

We used the double Gaussian deconvolution and the  diagnostic method 
suggested by Schneider and Young (1980) and  Shafter (1985) to measure radial 
velocity variations. Two Gaussians with fixed width and separation are 
convolved with an emission line. The position of equal intensity Gaussians 
corresponds to the line center. The width and separation could be changed 
in order to get use of different parts of line wings. Shafter (1985) describes 
the technique how to choose the optimal separation. We used a 4 \AA\, FWHM 
gaussians corresponding to our spectral resolution and follow the mentioned 
diagnostic method to pick up a 10 \AA\ half-separation as the best parameter 
to measure the emission line centers. 

\begin{figure*}
    \vbox{\psfig{figure=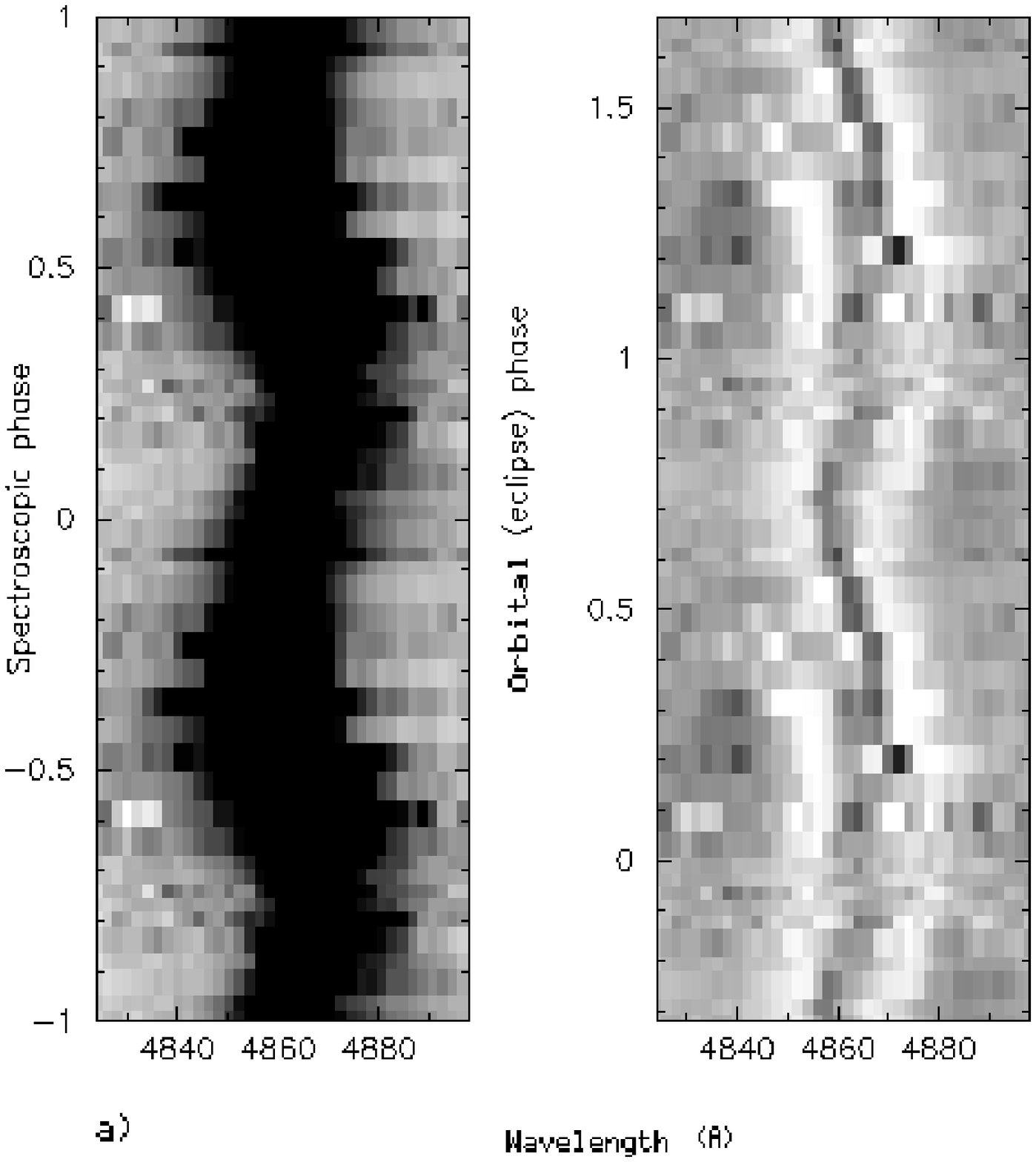,width=8.9cm,%
          bbllx=0.5cm,bblly=3.5cm,bburx=18.cm,bbury=22.2cm,clip=}}\par
     \vspace*{-9.5cm}\hspace*{9.2cm}
    \vbox{\psfig{figure=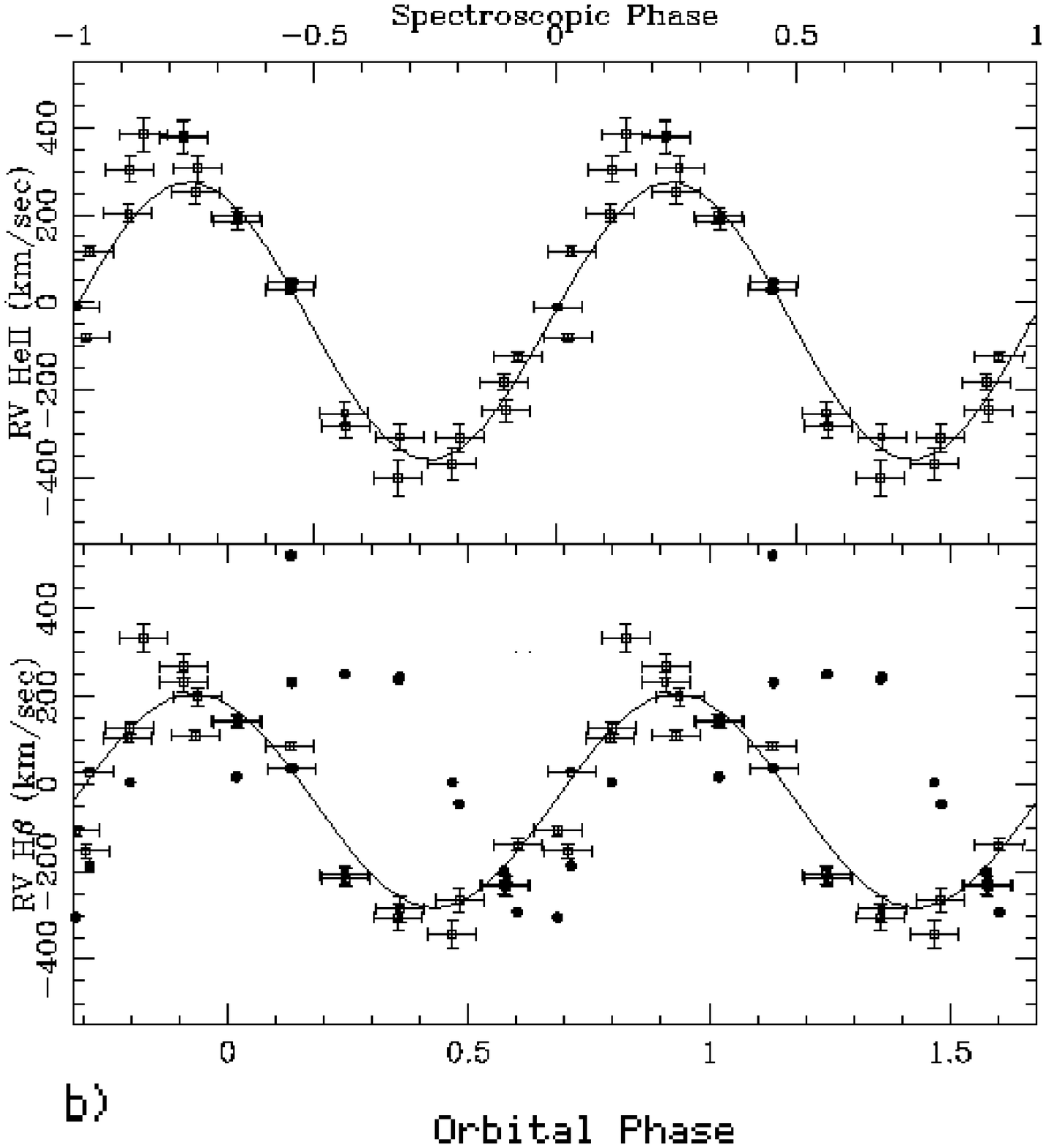,width=8.5cm,%
          bbllx=1.4cm,bblly=4.8cm,bburx=20.0cm,bbury=25.50cm,clip=}}\par
 \caption[rvcurve]{{\bf a)} The trailed spectra around H$\beta$. In the left 
panel the line wings are highlighted. In the right panel the same line 
is shown after subtraction of a gaussian fit with the center calculated from 
the spectroscopic phase.
{\bf b)} The radial velocity curves of He~II (top) and 
 H${\beta}$ (bottom). The error of measurements is marked by a vertical bar.
The exposure length is marked by a horizontal bar.  
 The solid circles in the lower panel indicate measurements of the emission 
feature as shown in the right panel of Fig. \ref{rv}a.
            }
\label{rv}
\end{figure*}

The spectroscopic period as derived from the line center variations of H$\beta$
shows maximum peak in the power spectrum at  $0.067916 \pm 0.000576$ days.
This coincides within reasonable limits 
with the 98.2 min.  period as derived from the moments of eclipses. 
The 98.2 min. (0\fd068207) period was therefore adopted as the orbital period 
of the system. The radial velocity (RV) curves (see Fig. \ref{rv}) were 
non-linearly fitted by the following function:

 \begin{equation}
V_{em}=\gamma + 
  K_{em}  \sin(2\pi\omega + \phi)
\label{rveq} 
       \end{equation}

\noindent where $\omega=t-T_0/P_{orb}$.

\noindent Phase zero (T$_0$) was chosen for H${\beta}$, so that $\phi=\pi n$, 
where n is an integer number. This implies the following ephemeris:

for H$\beta$: ~~$\gamma= -38.4$~km/sec\\
 \hspace*{1.65cm} T$_0=2450180.574398\pm0.00007$ \\
 \hspace*{1.38cm}   K$_{em} = 245.7\pm1.5$~km/sec

for He~II:  $ \gamma= -42.0$~km/sec\\
\hspace*{1.75cm}  T$_0=2450180.57348\pm0.00009$ \\
\hspace*{1.48cm} K$_{em} = 318.4\pm2.3$~km/sec

According to this, the --/+ crossing of the RV curve occurs at orbital phase 
$\phi_{orb}=0.68$ (for H$\beta$). The obtained amplitudes of the radial 
velocities are significantly higher than one
 would expect from orbital motion of the WD, but would not be unexpected for 
emission formed in an accretion stream, where the intrinsic velocities can be 
quite high. 

Although we tried a wide range of gaussian separations, the procedure was 
uncapable to distinguish the presence of components other than measured in the
emission lines. However, a careful study of the trailed spectra shows complex 
structure of emission lines during some phase intervals. In order to reveal 
the weaker component we fitted H$\beta$ with a single gaussian with the center 
calculated from  Eq. \ref{rveq} and variable widths. Then we subtracted the 
fit from the actual line profile. The residuals, in form of trailed spectra, 
are presented in the right panel of Fig. \ref{rv}, while on the left panel 
the  original spectra (normalized to the continuum) are displayed with levels 
of grey allowing the best visualization of the spectroscopic period.
The residual spectra bear inside a weak trace of emission shifted in phase 
to the spectroscopic period. The feature varies in strength and is most 
prominent at orbital phases $ 0.25>\phi_{orb}>0.75$ marked on the right panel. 
We measured the radial velocity variations of this feature and display these 
in the lower panel of Fig. \ref{rv}  (solid circles). Considering its 
period to be equal to the orbital one, and assigning the bad shape to our 
inaccuracy of the measurements, we assume that its --/+ crossing occurs 
at the eclipse and that it becomes brighter
 at phases when we are facing the secondary component.

\begin{figure*}
      \vbox{\psfig{figure=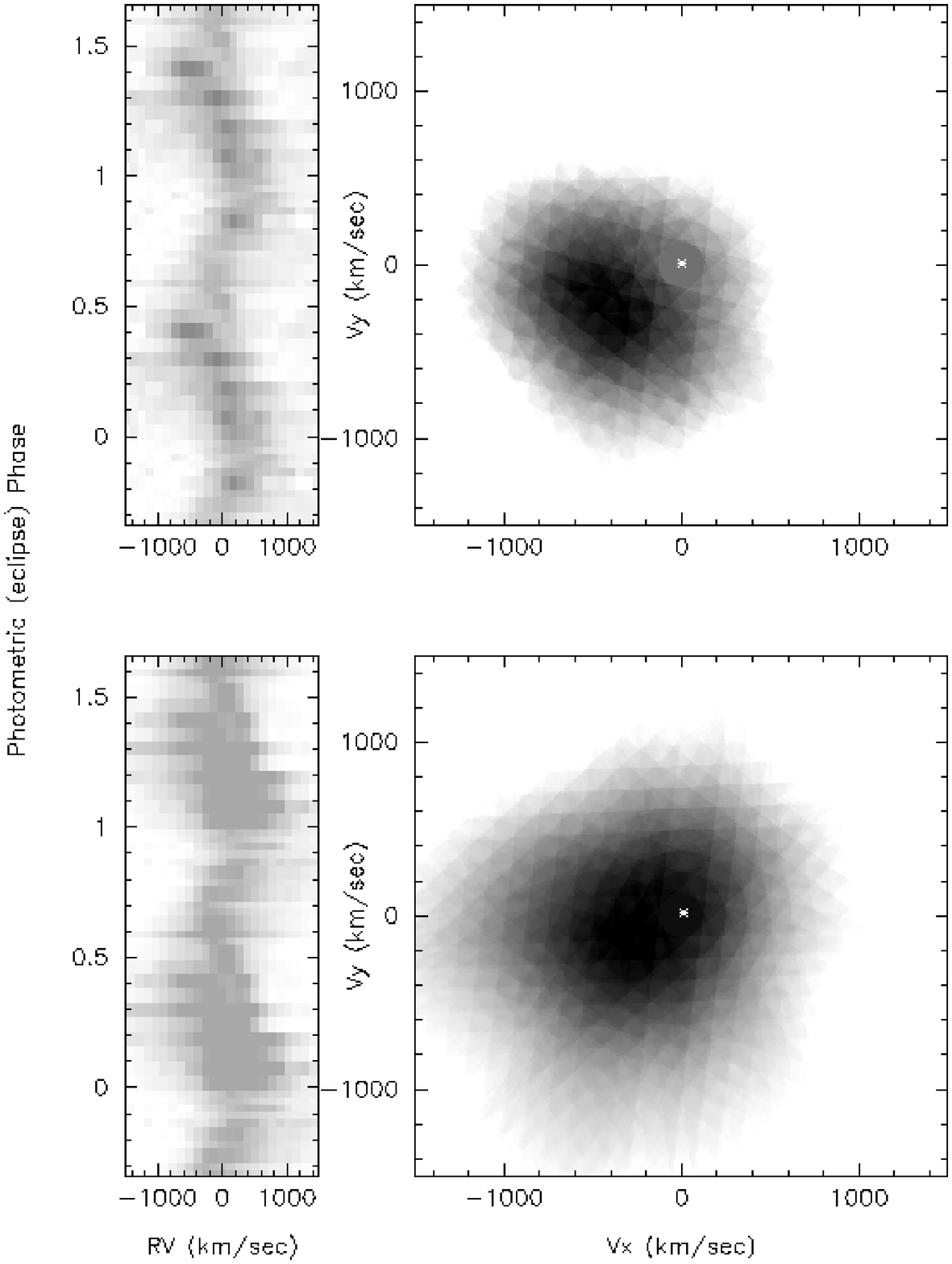,width=16.5cm,%
          bbllx=0.5cm,bblly=1.1cm,bburx=21.6cm,bbury=27.3cm,clip=}}\par
 \vspace{-0.25cm}
 \caption[tomo]{The trailed spectra and images of the velocity plane of \rxj\, 
        of  H${\beta}$ (bottom) and He~II (top) obtained by the backprojection
        method. 
        The center of mass of the system is marked on the $V_x-V_y$ images
        by a white symbol.} 
\label{tm}
\end{figure*}

The filtered backprojection method (Marsh \& Horne 1988) was applied to the 
lines of HeII and 
H${\beta}$ with the orbital (photometric) phase registration adopted above. 
As we noted above the eclipse probably does not coincide precisely with the
conjunction of the stellar components. However, its phase shift is not 
significant for the figure we obtain by doppler tomography since 
it gives us a general impression of the system configuration, but no 
quantitative measurements. The resulting velocity 
maps (tomograms) are presented in Fig. \ref{tm}. The spectra of both 
lines are displayed as trailed gray-scale images on the left.  
On the maps the
secondary is located along the $+V_Y$ axis and the WD along $-V_Y$. It is 
evident that the He~II emission region is compact and concentrated at the 
expected location of the accretion stream/hot spot. Unlike He~II, 
H${\beta}$ shows a more diffuse distribution 
and spreads out to  the inner $L_1$ point. 

\subsection{Geometric Configuration}

Given the presence of eclipses, the inclination $i$ of the orbital plane
of the system is 78\degs$\lax i \le$ 90\degs. The accretion stream with its 
hot spot is the main contributor of radiation. It is the hot spot which we 
see being occulted
and it is also the major contributor of the emission lines, producing radial 
velocity variations responsible for spectroscopic phasing. The other 
observable component in the emission lines is due the secondary with the 
heated surface facing the hot, bright spot. The irradiated part of the 
secondary produces the single humped light curve in the R band and the weak 
component of emission lines. The hot spot is located on the accretion 
stream but it is not yet clear where it is along  the stream. The possible 
locations are: during the ballistic part before the coupling to the WD 
magnetic field, i.e. when the stream is still inside the orbital plane
or shortly before the shock, i.e. about 0.1 WD radii above the WD surface.
The latter would be the more obvious location, but there we would expect
much higher velocities, which we do not see.

\subsection{The red and near-infrared spectrum}

\begin{figure}
      \vbox{\psfig{figure=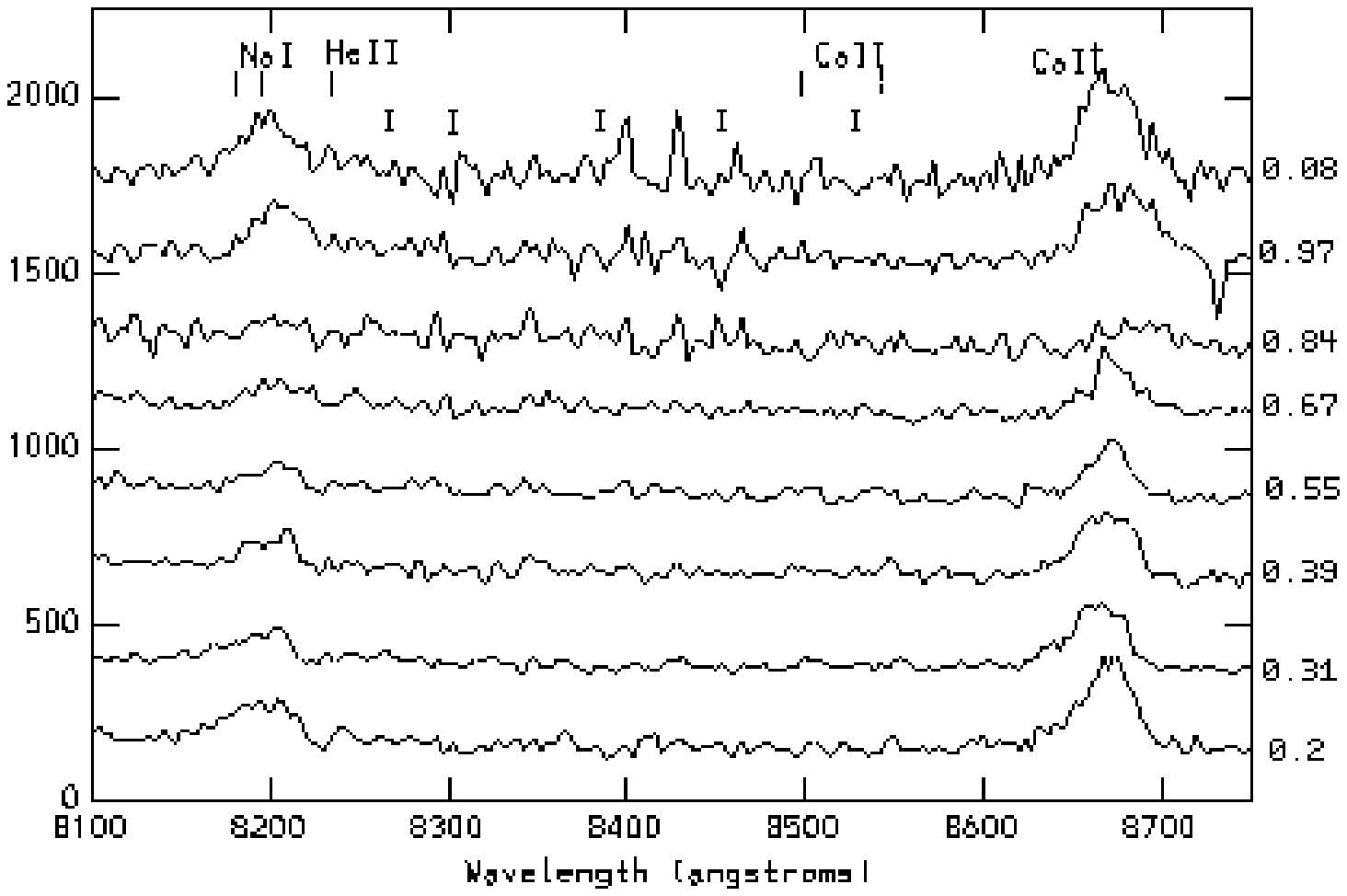,width=8.8cm,%
          bbllx=3.3cm,bblly=8.9cm,bburx=18.4cm,bbury=19.2cm,clip=}}\par
 \caption[naca]{The evolution of spectra in the spectral region between 
  $\lambda\lambda 7800-8900 \AA$ through different orbital phases as marked on 
  the right hand side. The numbers on the left side of the panel refer to the 
  counts of the bottom spectrum. Other spectra are shifted up by 100\% each.
  The spectral lines usually present in CV spectra are marked on the top. 
  Vertical ticks mark some prominent TiO features in the spectra of M4--M6 
  red dwarfs.}
\label{naca}
\end{figure}

The near-infrared spectra reveal two strong emission features at 
$\lambda~8200~\AA$ and $\lambda~8660~\AA$ (Fig. \ref{naca}). These coincide 
with the location of the Na~I doublet and one of the components of the 
Ca~II triplet. These lines could be recognised at some phases 
as absorption features seated on tremendously broad emission.
In a huge sample of 
CVs observed by Friend \etal (1988), no object was found to display such a 
behaviour, although similar but much fainter features showed up in some 
accretion disc systems after subtraction of the secondary's spectra. These 
features were referred to as ``disc component" by Friend \etal (1990). 
In our  case, however, these lines become prominent during eclipse and fade 
to halt at phase when we are looking along the accretion stream 
($\phi_{orb}=0.7 - 0.85 $).

At $\lambda~8204~\AA$ happens to be the Paschen jump. Mukai and Charles 
(1987) have also observed excess emission at that wavelength in PG~1550+191, 
a well known polar. They argue that it could be due to the peak of the 
residual 
Paschen continuum (blend of discrete lines at the limit of the Paschen series 
broadened by the Stark effect). However, they observed also lower series lines 
at significant strength, which is not the case for \rxj.

We also looked for possible absorption features in the red portion of the 
spectra, aimed in identifying the secondary star. 
The combined sensitivity of the spectrograph and CCD matrix drops rapidly 
redwards of
$\lambda~7500~\AA$. Although we are able to see strong emission features and 
their changes, the continuum and absorption features are below S/N=10 level,
and are only marginally distinguished. Unfortunately, we have not observed 
any red dwarf standards simultaneously for comparison and cross-correlation.
 
We could not identify the strong TiO band features one would expect if we had 
a significant contribution of the secondary in the red part of the spectrum.
Nevertheless, we found some hints witnessing a M~4--M~5.5 spectral type of 
the secondary. Young \& Schneider (1981) and later Wade \& Horne (1988) 
found quantitative spectral indices to estimate the spectral types of CV 
secondaries. Equivalent widths of the absorption features cannot be used 
directly in most cases, because they are affected by continuum from the 
WD+accretion structure, but flux deficits relative to a reference 
continuum remain unaffected.
We followed roughly the procedure described by Wade \& Horne (1988). First, we 
subtracted the telluric line at $\lambda~7600~\AA$ by fitting each 
component by a gaussian using the outer wings on three spectra obtained in 
phases 0.9--0.1. Then we fitted the continuum to the spectra with a second 
order polynom, which is almost a straight line in that part of the spectrum, 
and finally calculated the flux deficits of $\lambda~7665/\lambda~7165~\AA$ of 
the TiO bands. The obtained values range between
$1.0 < d_{\nu}(\lambda~7665)/d_{\nu}(\lambda~7165) < 1.5$, which corresponds 
to a M4--M6 spectral type. The accuracy of our measurements does not allow 
a more precise estimate.  This is not 
surprising, because the majority of the short-period  CVs below the period gap 
have late-type companions (Echevarr\'{\i}a (1983), Tovmassian (1984)).

The absolute magnitude of a M5-6 main-sequence star (the non-illuminated side)
ranges between M$_{\rm V}$ = 12.3--13.5 mag. During the eclipse - when we
see the back-side of the companion - the V brightness drops down to 
V$\approx$19 mag. This implies a lower limit for a distance of 100--150 pc, 
consistent 
with the rough guess from the X-ray absorption measure. 

\section{Summary}

The basic results of our observational study can be summarized as follows:
   \begin{enumerate}
      \item Based on the nature of the X-ray spectrum, the synchronized WD 
spin and orbital period, and despite the moderate strength of He~II relative 
to H${\beta}$, the ROSAT all-sky survey source \rxj\, is identified as a new
eclipsing polar. The orbital period is found to be 98.2 min. Narrow 
eclipses are observed which correspond to the occultation of the hot spot 
on the accretion stream by the companion. The eclipse period coincides 
with the radial velocity variations of the major emission lines.
      \item The high amplitude of the radial velocity variations and its 
phasing relative to the eclipses allow us to conclude that the major line 
emission also comes from the frame of the accretion stream and hot spot. 
However, line emission requires different conditions (e.g. density)
then continuums emission, so that the two emission regions are expected
to be distinct. We were successful to retrieve also the weaker component 
of emission lines formed on the heated side of the secondary. The geometric 
interpretation  fits well the usual picture of eclipsing polars in  analogy 
with other similar magnetic cataclysmic variables. 
      \item From the absorption of the X-ray spectrum and the inferred
absolute magnitude of the companion as compared to the observed one we
estimate a lower limit of the distance to d $\lax$ 100--150 pc.  
The phase-averaged, unabsorbed X-ray luminosity is 
2.5$\times$10$^{30}$ (D/100 pc)$^2$ erg/s in the 0.1--2.4 keV range.
    \item The Doppler tomography localizes the source of strong He~II line
emission to the accretion stream/hot spot. Other lines with lower excitation 
levels are more spread out over the system. 
   \item The detection of the two emission lines at 8200 \AA\ and 8660 \AA\ is 
one of the most exciting (and barely understood) results of these observations.
Although we were unable 
to identify them with any known lines, their presence implies the existence 
of conditions in \rxj\, which arise in very few systems only (based on the lack
of detections in many other polars). Unfortunately, the 
amount of existing data does not yet allow any further conclusions.
\item A low magnetic field for the white dwarf is suggested by three facts:
(a) The optical spectrum does not show hints for cyclotron lines.
(b) The X-ray spectrum is observed to be rather hard. The 
soft-to-hard X-ray flux ratio for polars with measured magnetic field strengths
has been shown to increase with magnetic field (Beuermann \& Schwope 1994).
(c) The stronger photometric modulation in red than in blue and its phasing also argues for a 
  low magnetic field.
   \end{enumerate}

\begin{acknowledgements}
      GT thanks Allen Shafter for discussing details of the reduction 
technique  and providing parts of the analysis code. The
authors are  grateful to R. Schwarz and to the referee M. Mouchet for valuable 
comments and discussions which considerably improved the presentation of our
results. This work was partially financed by the DGAPA project IN109195.
JG is supported by the Deutsche Agentur f\"ur
Raumfahrtangelegenheiten (DARA) GmbH under contract FKZ 50 OR 9201
and 50 QQ 9602 3.
The \ros\, project is supported by the German Bundes\-mini\-ste\-rium f\"ur
Bildung, Wissenschaft, Forschung und Technologie (BMBF/DARA) and the
Max-Planck-Society.
\end{acknowledgements}

\end{document}